\documentclass[a4paper,11pt]{article}
\pdfoutput=1 % if your are submitting a pdflatex (i.e. if you have
\usepackage{jinstpub} % for details on the use of the package, please
\title{Characterisation of novel thin n-in-p planar pixel modules for the ATLAS Inner Tracker upgrade}

\author[1]{J.-C.~Beyer,\note{Corresponding author.}}
\author{A.~La~Rosa,}
\author{A.~Macchiolo,}
\author{R.~Nisius,}
\author{N.~Savic}
\author{and R.~Taibah}

\affiliation{Max-Planck-Institut f{\"u}r Physik (Werner-Heisenberg-Institut), \\ F{\"o}hringer Ring 6, DE-80805 M{\"u}nchen, Germany}

\emailAdd{jbeyer@mpp.mpg.de}

\abstract{In view of the high luminosity phase of the LHC (HL-LHC) to start operation around 2026, a major upgrade of the tracker system for the ATLAS experiment is in preparation. The expected neutron equivalent fluence of up to $2.4\times10^{16}\,$1 MeV~n$_\text{eq.}/$cm$^2$ at the innermost layer of the pixel detector poses the most severe challenge. Thanks to their low material budget and high charge collection efficiency after irradiation, modules made of thin planar pixel sensors are promising candidates to instrument these layers. To optimise the sensor layout for the decreased pixel cell size of $50\times50\,\mu\mathrm{m}^2$, TCAD device simulations are being performed to investigate the charge collection efficiency before and after irradiation. In addition, sensors of 100-$150\,\mu\mathrm{m}$ thickness, interconnected to FE-I4 read-out chips featuring the previous generation pixel cell size of $50\times250\,\mu\mathrm{m}^2$, are characterised with testbeams at the CERN-SPS and DESY facilities. The performance of sensors with various designs, irradiated up to a fluence of $1\times10^{16}\,$n$_\text{eq.}/$cm$^2$, is compared in terms of charge collection and hit efficiency. A replacement of the two innermost pixel layers is foreseen during the lifetime of HL-LHC. The replacement will require several months of intervention, during which the remaining detector modules cannot be cooled. They are kept at room temperature, thus inducing an annealing. The performance of irradiated modules will be investigated with testbeam campaigns and the method of accelerated annealing at higher temperatures.}


\proceeding{11$^{\text{th}}$ Workshop on Position Sensitive Detectors\\
  3$^{\text{rd}}$-8$^{\text{th}}$.08.2017\\
  The Open University, Milton Keynes, England}

\begin{document}
\maketitle
\flushbottom

\section{Introduction}

The Large Hadron Collider (LHC) is facing its upgrade to achieve an increase of the instantaneous luminosity of about a factor of five. Along with the increase in instantaneous luminosity, the track density in the tracking system will increase, making smaller pixel cells essential to keep the present occupancy levels and tracking resolution unchanged. The instantaneous luminosity of $5-7\times10^{34}\,$cm$^{-2}$s$^{-1}$ will integrate over ten years of running to around $3000-4000\,$fb$^{-1}$ leading to a to a $1\,$MeV neutron equivalent fluence fluence of $2-2.4\times10^{16}\,$n$_\text{eq.}/$cm$^2$ at the innermost parts of the experiments \cite{BSmart,StripsTDR}.

The ATLAS detector will be equipped with a new Inner Tracker (ITk) to cope with the harsher conditions. It features an all silicon layout consisting of 5 pixel layers. To maintain the current occupancy levels, the pixel cell size is reduced from the present $50\times400\,\mu\mathrm{m}^2$ (layer 1-3, assembled with the FE-I3 readout chip) and $50\times250\,\mu\mathrm{m}^2$ (layer 0, Insertable B-Layer, assembled with the FE-I4 readout chip \cite{FEI4}) to $50\times50\,\mu\mathrm{m}^2$ or $25\times100\,\mu\mathrm{m}^2$. The necessary new read-out chip is currently under development by the RD53 collaboration in the TSMC $65\,$nm CMOS technology \cite{RD53}. Due to the harsh radiation environment in the innermost layers, an exchange of the two innermost layers is foreseen after half of the detectors' lifetime, thus, reducing the maximal fluence to $1-1.2\times10^{16}\,$n$_\text{eq.}/$cm$^2$ for the innermost layer. The replacement of these two layers may cause an interruption of the pixel detector cooling system. The third layer, having received a fluence in the range of $1-2\times10^{15}\,$n$_\text{eq.}/$cm$^2$ at this point in time, will undergo a reverse annealing, leading to an increase in full depletion voltage and a decrease in charge collection.

This paper presents Technology Computer Aided Design (TCAD) device simulation of thin planar pixel sensors which are a candidate to instrument all pixel layers, given their low material budget, high charge collection efficiency after irradiation and low power dissipation. The simulation is carried out to optimise the pixel implant size to achieve the highest charge collection before and after irradiation, within the constraints given by the $50\times50\,\mu\mathrm{m}^2$ pitch of the RD53 chip. In addition, the dependence of the breakdown voltage on the p-spray doping is studied. Finally, a testbeam based study of the efficiency of annealed irradiated pixel modules is presented.

\section{Thin planar pixel sensors}

\begin{figure}[htbp]
\centering
\includegraphics[width=1\textwidth]{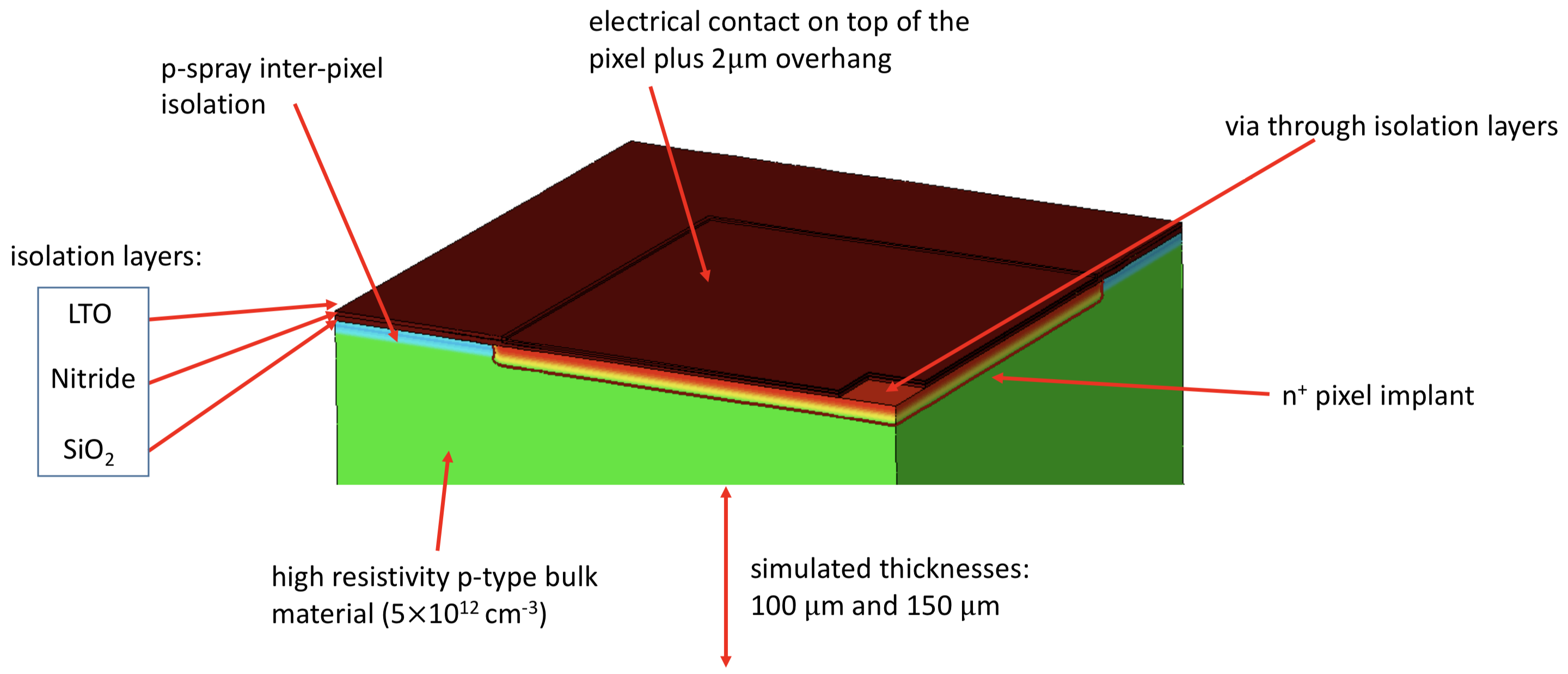}
\caption{\label{fig:pixelschematic} Schematic of the simulated pixel showing various aspects of the design and technology.}
\end{figure}

Thin planar pixel sensors with a thickness of $100\,\mu\mathrm{m}$ have been shown to be working after fluences ranging from 1 to $10\times10^{15}\,$n$_\text{eq.}/$cm$^2$ while fulfilling the requirements to instrument layer 1-4 of the ITk \cite{NSavic}. The moderate bias voltage necessary to obtain a tracking efficiency $\geq97\%$ and their lower power dissipation compared to thicker sensors make them particularly suited for the application at the fluence levels of the high-luminosity upgrade of the LHC (HL-LHC). A schematic of the simulated pixel is shown in Fig. \ref{fig:pixelschematic}.

They are based on n-in-p technology with a homogeneous maskless low-dose p-spray boron implantation. The p-spray compensates the electron-attracting positive Si-SiO$_2$ interface charge which otherwise would lead to a conductive near-surface layer shorting all pixel implants. One of the key parameters determining the breakdown voltage are the p-spray properties. Since at a given applied potential, the electric field in the bulk is proportional to the gradient of the doping concentration, it is maximal at the junction of n$^+$ pixel implants and shallow p-spray implantation. To improve the breakdown behaviour of sensors before irradiation, the influence of the p-spray doping profile with respect to the breakdown voltage is analysed. Investigations are performed with simulations and measurements on a pixel production carried out with different p-spray implant parameters at the Semiconductor Laboratory (HLL) of the Max-Planck Society. To process very thin substrates, 6" Silicon on Insulator (SOI) wafers were used with $100\,\mu\mathrm{m}$ and $150\,\mu\mathrm{m}$ active thickness. After the post-processing phase where the Under Bump Metal (UBM) is deposited, the technology foresees the removal of the backside handle wafer, which is required for mechanical stability of the thin silicon wafers during all processing steps. Besides sensors from this HLL production, also FE-I4 modules with sensors from a previous SOI production at HLL as well as modules with sensors from a 4" wafer production at CiS \cite{CiS} were tested.

\section{Simulation methods and setup}

%tcad synopsys version, describe 3D model incl layers etc., pspray depth peak and concentration, backside electrode, bulk doping, irradiation via traps and surface, traps pennicard, surface 1e12 (also in pennicard), Temperature, bias voltages, boundary conditions, bandgap model, mobility: dopingdep, recomb: SRH, Auger, hAvalanche, HION: 80e+- per micron, 1um gaus, CCE: integrate 20ns and subtract baseline

For all presented simulations, the finite-element semiconductor simulation package Synopsys Sentaurus TCAD version L-2016.03 was used. A 3D model of the pixel sensor is approximated by a mesh structure of individual finite-elements (vertices). To reduce the necessary computational effort, the structure can be reduced down to the smallest symmetry, in this case a quarter of a pixel. The continuation of the structure is dealt with Neumann boundary conditions requiring the electrical fields and current densities to be zero normal to the boundaries of the mesh. This requirement and hence, the high accuracy of the reduced structure is given if planes of symmetry are chosen as boundaries. Reference \cite{Rainer} has shown that this assumption is valid for clean wafers before irradiation.
\\
The simulated structure is one quarter of a square pixel, leading to a simulation volume of $25\times25\times\mathrm{thickness}\,\mu\mathrm{m}^3$, defined by the RD53 pixel pitch of $50\times50\,\mu\mathrm{m}^2$. It includes the insulation layers of high temperature oxide, nitride and Low Temperature Oxide (LTO) on top of the silicon bulk, to reproduce the actual process of the manufactured sensors. An electrode mimicking the aluminium on the real pixel overlaps the pixel implant size by $2\,\mu\mathrm{m}$. The connection of the electrode to the pixel implant is realised through a $5\times5\,\mu\mathrm{m}^2$ via in the insulation layers. The bulk doping is about $1\times10^{12}\,$cm$^{-3}$ and the p-spray has a peak concentration of $6\times10^{16}\,$cm$^{-3}$ while it reaches $0.6\,\mu\mathrm{m}$ deep until the concentration arrives back to the nominal bulk value. The backside of the sensor has a homogeneous p$^+$ doping with an overlaying electrical contact. The bias voltage is applied to the sensor by keeping the pixel contact at ground while ramping the potential of the backside contact to (negative) high voltage.
\\
One important goal of this study is to investigate the consequences of irradiation depending on the sensor design. Irradiation causes two main categories of radiation damages: bulk damages and surface effects. Bulk damages manifest as an increase of leakage current, full depletion voltage and trapping probability leading to decreased carrier lifetime. This can be explained by the non-ionising-energy-loss (NIEL) hypothesis. Particles of high energy traverse the silicon sensor and loose energy by ionisation and scattering with bulk atoms. In case of significant energy transfer to the bulk atom (the threshold is at $21\,$eV \cite{Akkerman}) it is displaced from its original lattice position causing a vacancy. Depending on the actual scattering, the energy level of the vacancy can either be of acceptor or donor type, but will always be situated in the band-gap of silicon. The acceptor type vacancies can trap electrons and thereby explain the increase of full depletion voltage (filled traps are negatively charged and increase the effective p-type doping) and increased trapping. Since the vacancies also act as carrier generation centers, they also cause the increased leakage current. In Synopsys TCAD, bulk damages are simulated by the introduction of acceptor and donor type traps. The model chosen for the following simulations is taken from \cite{delhi} and foresees one acceptor and one donor type trap. This model was chosen as one amongst several available models and a comparison is planned for the future. While the differences in terms of absolute values between the available models might be significant, the comparison of the relative performance of different pixel geometries should not be influenced.
\\
The second category of radiation damage are surface effects. Ionising radiation releases electron hole pairs in the insulating SiO$_2$ layer grown on top of the bulk. Only the electrons have a considerable mobility whereas holes can only be transported via shallow traps in the oxide. Eventually, they are captured by deep traps forming a fixed positive space charge. Following the measurements of \cite{Zhang2012}, the oxide charge saturates after high ionising doses in the order of several hundred MGy at values of $\approx2-3\times10^{12}$cm$^{-2}$. Due to arising uncertainties from the production technology and given the possibly not completely saturated and already partially annealed oxide charge, the oxide charge in the simulation is fixed to $2\times10^{12}\,$cm$^{-2}$ for all irradiations and $5\times10^{10}\,$cm$^{-2}$ before irradiation.
\\
The temperature is set to $300\,$K for all un-irradiated devices and $254\,$K for all irradiated devices. The standard bias voltage is $50\,$V in the un-irradiated case and $500\,$V for irradiated structures. The charge collection efficiencies are obtained by simulating the impact of a minimum ionising particle (MIP) releasing 76 e/h pairs per $\mu\mathrm{m}$ with a gaussian width of $1\,\mu\mathrm{m}$. The transient of the device is simulated in these cases from $5\,$ns before the impact to $30\,$ns after the impact. The collected charge is obtained by integrating the current pulse and subtracting the baseline leakage current which is extracted from the $5\,$ns before the impact.

\section{Simulation of \texorpdfstring{\boldmath{$50\times50\,\mu\mathrm{m}^2$}}{50x50um2} pixel cells}

%BD vs p-spray (process optimizations, coming from moderated p-spray...), BD vs implant size (p-spray potential decreases with increasing implant size, plot for this?!) , CCE vs implant size (will increase but counteracted by noise coming from increased capacity, need bigger structure for this, thus not here), comparison to SOI4 IV?!

\subsection{Investigation of p-spray properties}

\begin{figure}[htbp]
\centering
\includegraphics[width=0.48\textwidth]{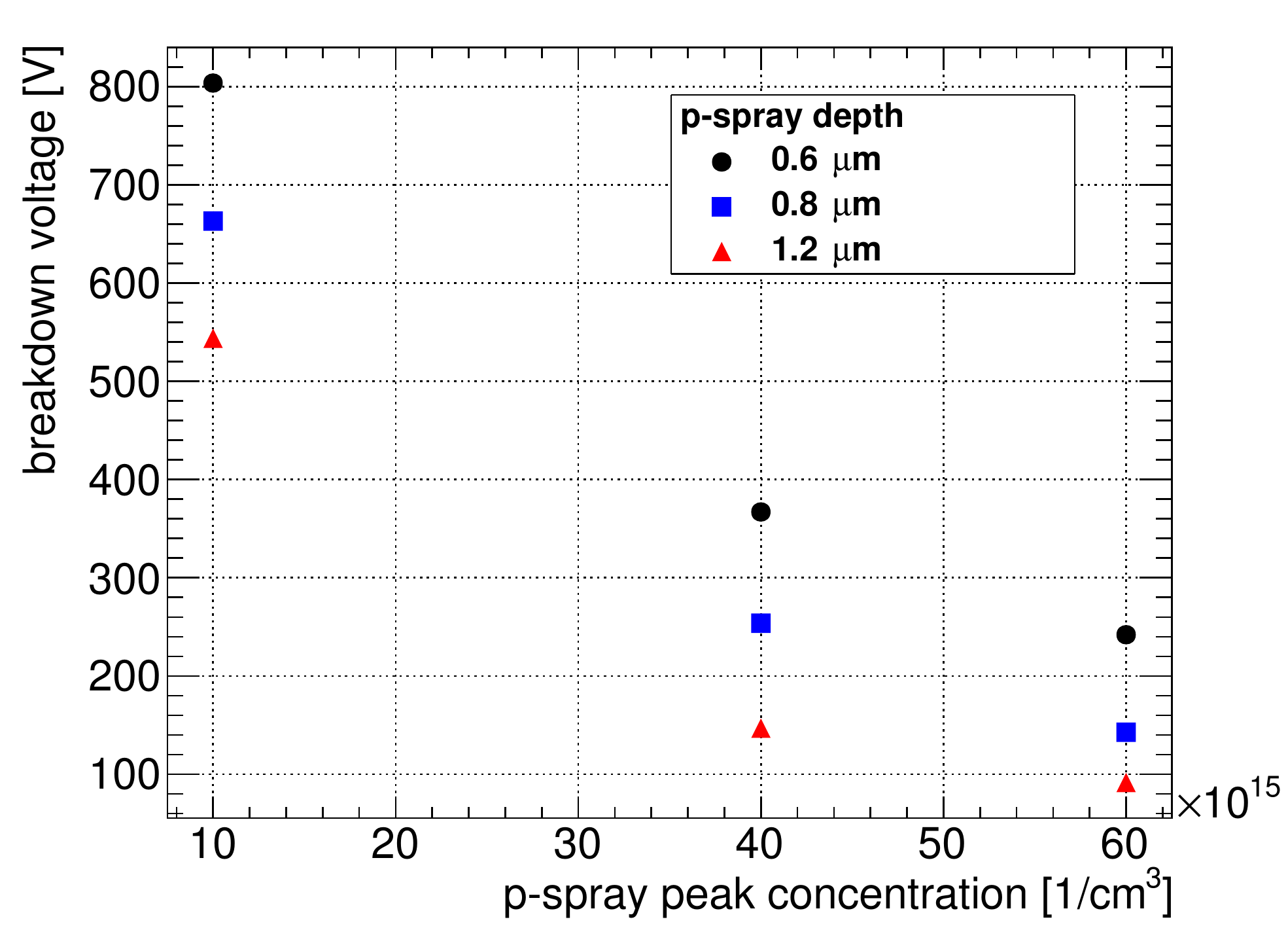}
\includegraphics[width=0.48\textwidth]{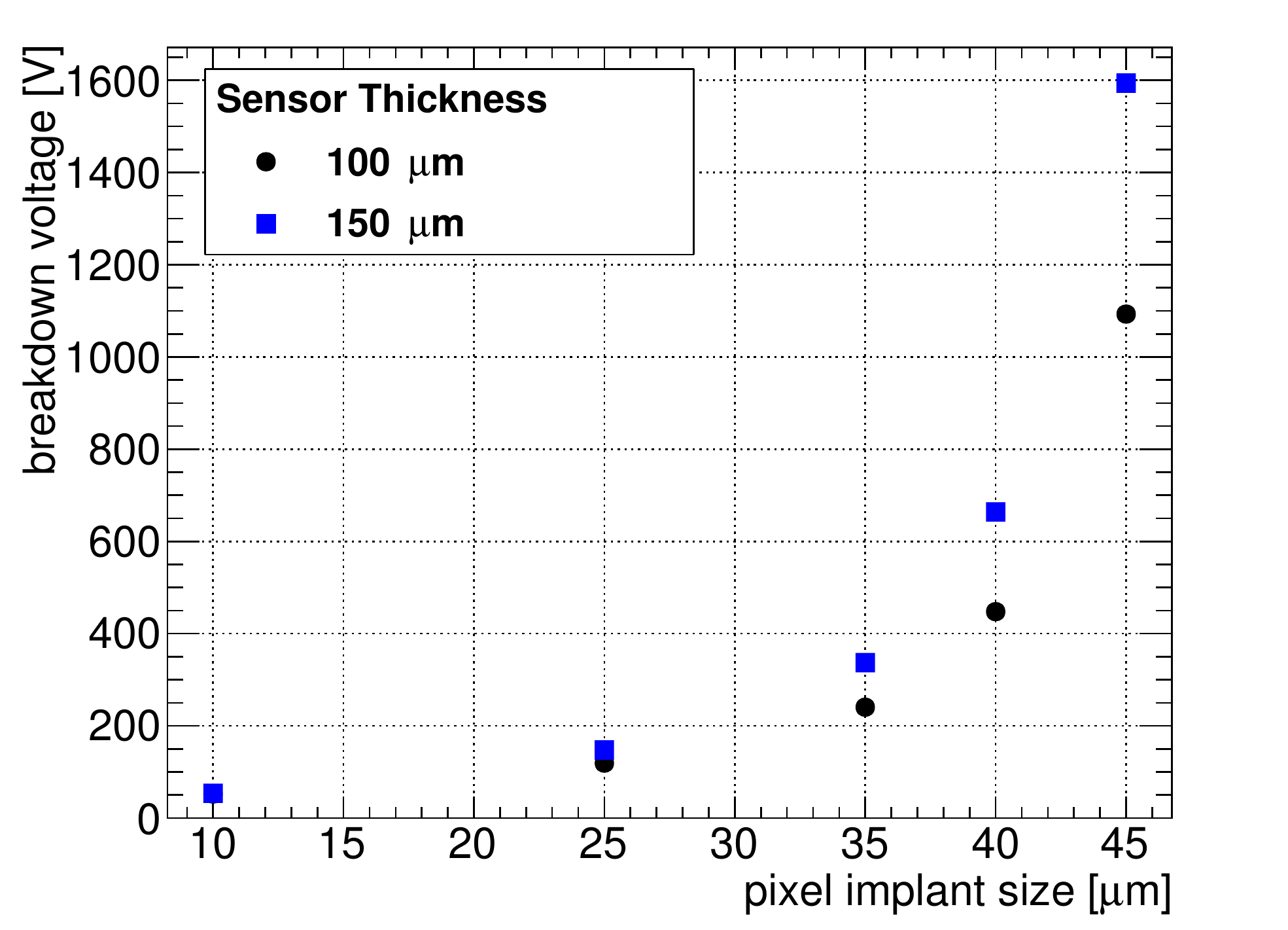}
\caption{\label{fig:bdvspspray_bdvsimplantsize} (Left) Simulated breakdown voltage as a function of the p-spray peak concentration for different p-spray depths for a pixel implant size of $35\,\mu\mathrm{m}$. (Right) Simulated breakdown voltage as a function of implant size for 100 and $150\,\mu\mathrm{m}$ thick sensors.}
\end{figure}

The breakdown mechanism within a pixel cell is determined by the highest field region which is defined by the highest gradient of the doping concentration. The highest gradient is given in the interface of p-spray and n$^+$ implant since the p-spray concentration is usually several orders of magnitude higher than the doping of the high resistive bulk material. The influence of two different p-spray parameters has been simulated: the depth of the implant (measured from the surface up to the point where the doping concentration returns to the bulk doping concentration) and the peak concentration. The implanted p-spray dose controls the peak concentration while the depth of the implant can be changed by the implantation energy and by the thermal treatment of the p-spray layer. Higher temperatures after the implantation lead to larger diffusion and deeper implants accompanied by a reduced peek concentration.
Figure \ref{fig:bdvspspray_bdvsimplantsize} (left) shows the simulated breakdown voltage of the pixel cell in the 3D model with a $35\,\mu\mathrm{m}$ pixel implant size for different p-spray depths as a function of the p-spray peak concentration. Lower peak concentrations and shallower implantations are beneficial with respect to the breakdown voltage. The breakdown voltage increases from highest to lowest peak concentration by about $450-550\,$V and from $1.2\,\mu\mathrm{m}$ to $0.6\,\mu\mathrm{m}$ p-spray depth by about $150-250\,$V.

\begin{figure}[htbp]
\centering
\includegraphics[width=0.7\textwidth]{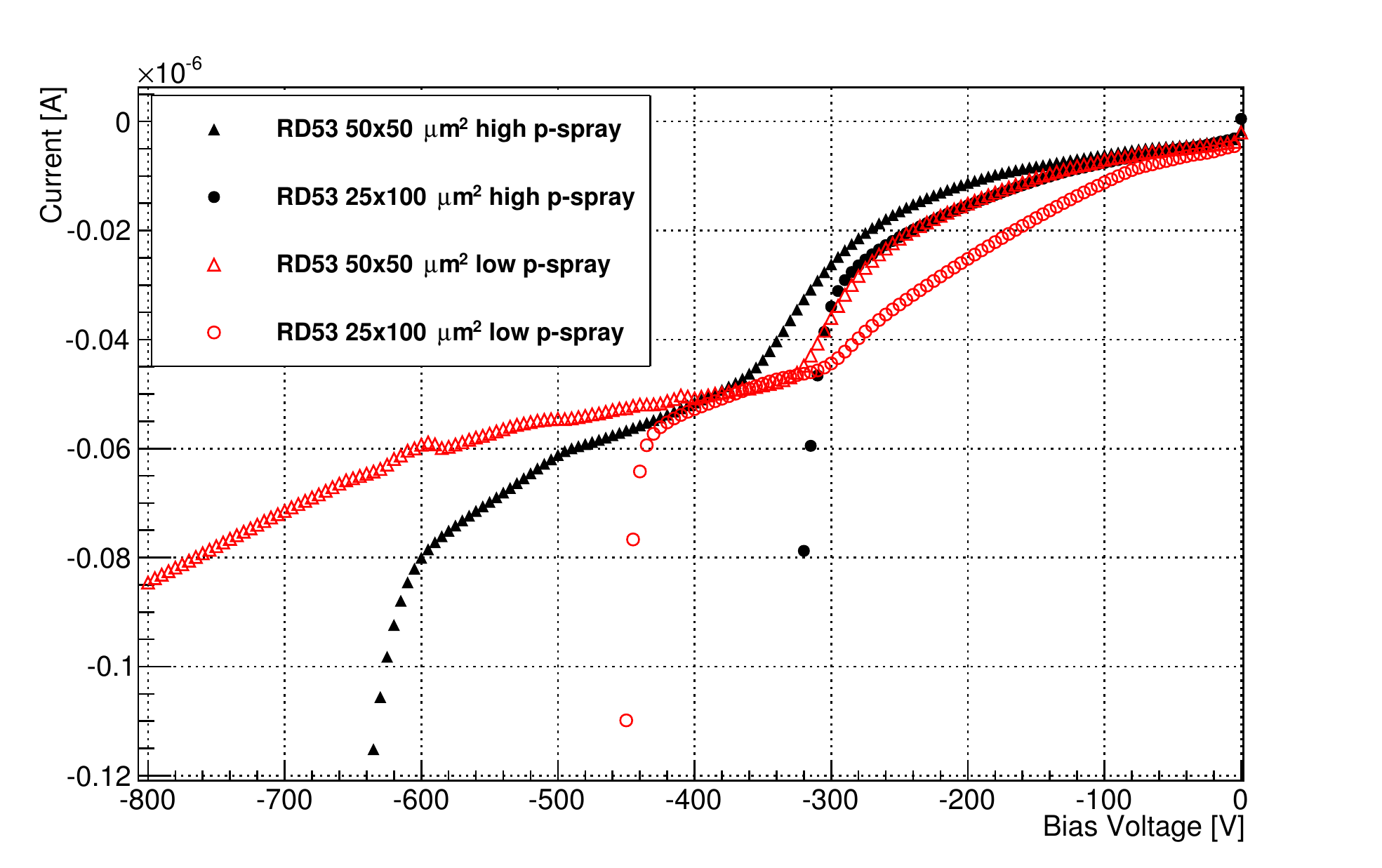}
\caption{\label{fig:ivsoi4} IV curves of two structures on two different wafers of the recent production at HLL. Both structures are sensors compatible to the RD53 read-out chip, featuring $50\times50\,\mu\mathrm{m}^2$ and $25\times100\,\mu\mathrm{m}^2$ pixel sizes. One wafer (black, filled symbols) was processed in the default way while the p-spray of the other wafer (red, open symbols) underwent a higher thermal treatment and a lower implanted dose resulting in a reduction of the peak doping concentration of the p-spray layer.}
\end{figure}

The findings have been tested in a recent production at HLL. The process of a $150\,\mu\mathrm{m}$ thick wafer was changed to have a lower p-spray implantation dose and an additional higher temperature treatment to reduce the peak concentration even if this results in a deeper diffusion of the implant. Figure \ref{fig:ivsoi4} shows the IV curves of two RD53 compatible ($50\times50\,\mu\mathrm{m}^2$ and $25\times100\,\mu\mathrm{m}^2$ pixel sizes) structures on two wafers of the recent HLL production. Since the exact geometries of the RD53 compatible sensors implemented in these wafers differ from the simulated structures they can not be directly compared to the simulated pixel cell \footnote{To test the sensors before interconnection to the read-out chip, in a fraction of the devices a punch-through structure is implemented which not only enables testing but also causes a different field configuration with respect to the simulated cell making direct comparisons difficult.}. Nevertheless, the overall trend of higher breakdown voltages in the low dose wafer (lower p-spray dose, higher thermal treatment) compared to the default process wafer (higher p-spray) agrees with the results of the simulation. IV curves of sensors without any biasing structure will allow for a more detailed comparison of simulation and data after the interconnection to read-out chips.

\subsection{Implant size optimisations}

\begin{figure}[htbp]
\centering
\includegraphics[width=0.7\textwidth]{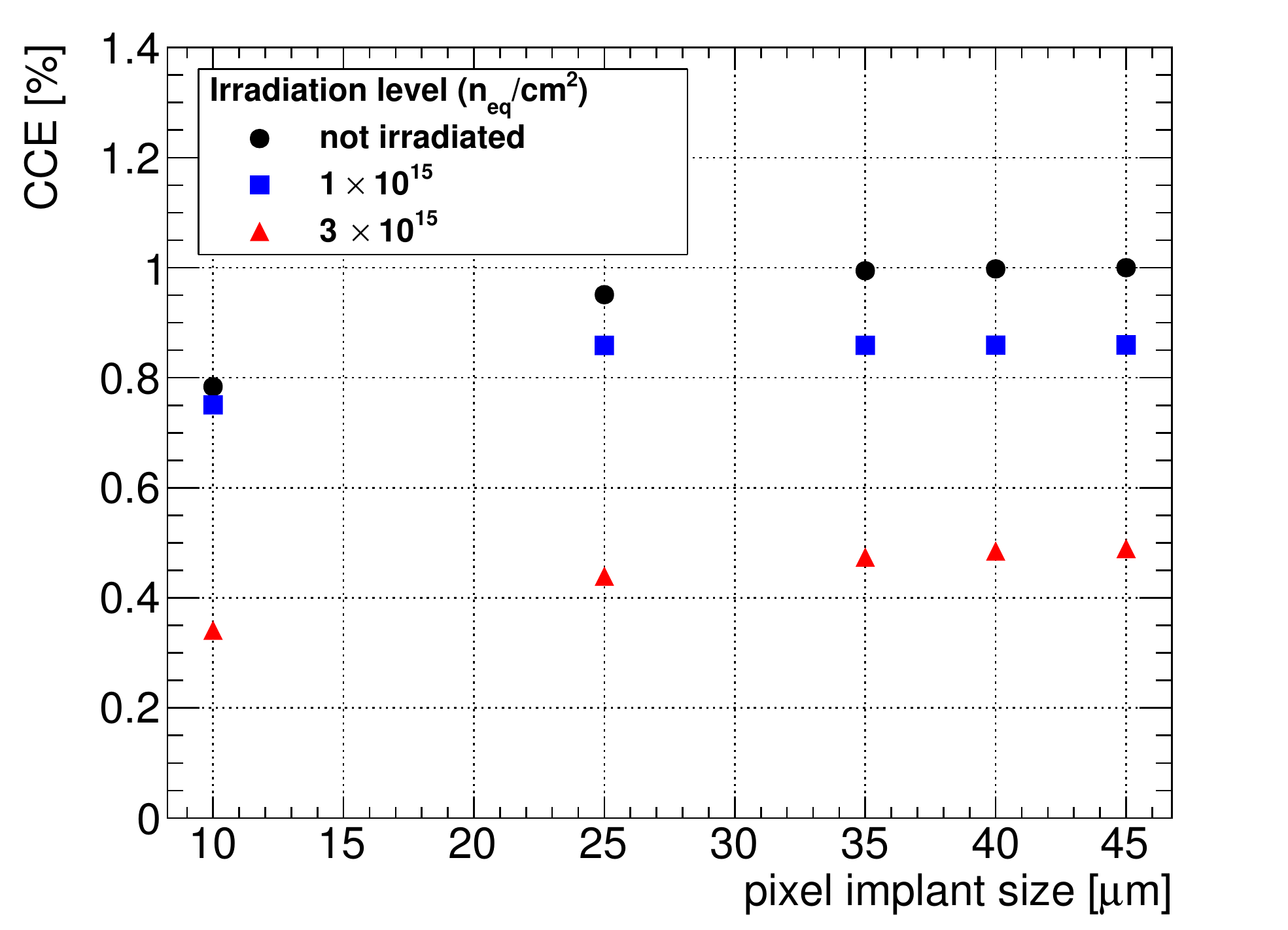}
\caption{\label{fig:CCEvsimplantsize} Charge collection efficiency before and after irradiation to a fluence of $1\times10^{15}\,$n$_\text{eq.}/$cm$^2$ and $3\times10^{15}\,$n$_\text{eq.}/$cm$^2$ as a function of the pixel implant size.}
\end{figure}

In addition to the process details described in the previous section, design parameters play a crucial role in the breakdown behaviour of a pixel sensor. This section will focus on the effect of different pixel implant sizes. Smaller implants lead to larger gaps in between the pixels. While the pixel implant is kept at ground potential, the p-spray potential is floating between the ground potential of the pixel implant and the negative potential of the backside contact. Smaller gaps between the pixel implants lead to an improved shielding of the p-spray against the backside potential. Consequently, it stays closer to the ground potential of the pixel implant compared to the case of larger gaps. A higher potential difference between pixel implants and p-spray layer causes higher fields in their interface region, causing an earlier breakdown. Figure \ref{fig:bdvspspray_bdvsimplantsize} (right) shows the simulation of the breakdown voltage as a function of implant size for sensor thicknesses of 100 and $150\,\mu\mathrm{m}$. An expected increase of breakdown voltage can be observed for increasing pixel implant size. The breakdown voltage of $150\,\mu\mathrm{m}$ thick sensors is higher compared to their $100\,\mu\mathrm{m}$ thick equivalents.
%\begin{figure}[htbp]
%\centering
%\includegraphics[width=0.6\textwidth]{embed_BDvsSize_100and150thickness_Nit.pdf}
%\caption{\label{fig:bdvsimplantsize} Simulated breakdown voltage as a function of implant size for 100 and $150\,\mu\mathrm{m}$ thick sensors.}
%\end{figure}
The influence of the pixel implant size on the charge collection efficiency (CCE) as another important benchmark is also investigated. Figure \ref{fig:CCEvsimplantsize} shows the CCE as a function of the pixel implant size before and after irradiation to a fluence of $1\times10^{15}\,$n$_\text{eq.}/$cm$^2$ and $3\times10^{15}\,$n$_\text{eq.}/$cm$^2$. The efficiency values are calculated relative to the highest collected charge. Before irradiation, the differences in collected charge for implant sizes of $35-45\,\mu\mathrm{m}$ are negligible. For $25\,\mu\mathrm{m}$ the collected charge already decreases by more than $5\%$ and for $10\,\mu\mathrm{m}$ it is only $78.4\%$ of the possible charge. After an irradiation to $1\times10^{15}\,$n$_\text{eq.}/$cm$^2$ the absolute collected charge decreases due to charge carrier trapping. The differences between different implant sizes get further reduced, being only $0.1\%$ difference between $25\,\mu\mathrm{m}$ and $45\,\mu\mathrm{m}$ implant size. The $10\,\mu\mathrm{m}$ large pixel collects still significantly less charge ($87\%$ of the $45\,\mu\mathrm{m}$ large implant). After an irradiation to $3\times10^{15}\,$n$_\text{eq.}/$cm$^2$, the collected charge of the implant sizes from $35\,\mu\mathrm{m}$ to $45\,\mu\mathrm{m}$ still only differ by $3\%$ while the $25\,\mu\mathrm{m}$ ($10\,\mu\mathrm{m}$) large pixel collects only $90\%$ ($70\%$) of the charge of the $45\,\mu\mathrm{m}$ large implant.

It is known that the inter-pixel capacitance increases with increasing pixel size. Since the capacity of the pixel determines the input noise of the readout channel, it could partially counteract the beneficial effects of an increasing CCE. A detailed study of the inter-pixel capacitance is planned for the near future.

\section{Testbeam studies of annealed pixel modules}

The microscopic defects in the silicon crystal lattice caused by NIEL damage are not stable in time. They undergo an annealing process \cite{mollthesis} which leads to the deactivation of defects on a short timescale, causing a decreased full depletion voltage and higher charge collection. However, on a longer timescale, more defects are activated than deactivated. An increase of full depletion voltage exceeding the situation before annealing is caused. The timescales of these beneficial and reverse annealing effects strongly depend on the ambient temperature. While both processes are strongly suppressed at the usual operational temperatures of modern pixel detectors well below $0^\circ$C, the timescale of beneficial annealing is in the order of one week and the timescale of reverse annealing becomes an order of several weeks at room temperature (RT).
\\ \\
To evaluate possible risks associated with cooling interruptions for the ITk pixel detector, a systematic study is carried out to measure the effect of reverse annealing on the hit efficiency of highly irradiated modules. Due to the absence of the new RD53 readout chip, modules consisting of thin planar n-in-p pixel sensors interconnected to FE-I4 chips of the IBL generation with a pixel pitch of $50\times250\,\mu\mathrm{m}^2$ are used. These modules were tested at the testbeam facilities of CERN-SPS and DESY. A detailed description of the methodology of testbeam measurements can be found in \cite{Testbeam}. One major cooling interruption is foreseen at the moment during the exchange of the two innermost pixel layers at half of the ITk lifetime when the outermost three layers will be exposed to room temperature conditions for several months.

\begin{figure}[htbp]
\centering
\includegraphics[height=5.8cm, width=0.42\textwidth]{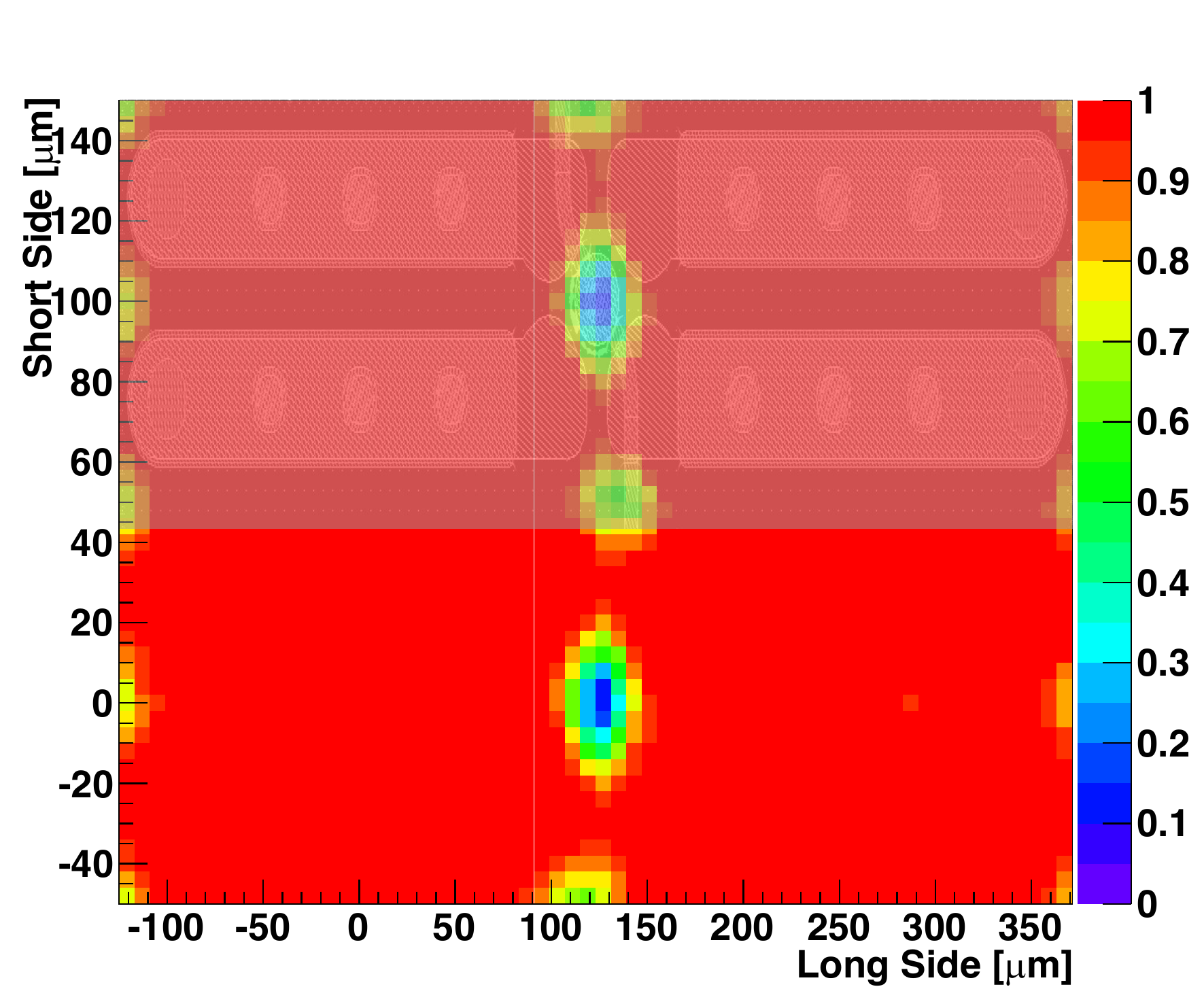}
\includegraphics[height=5.8cm, width=0.42\textwidth]{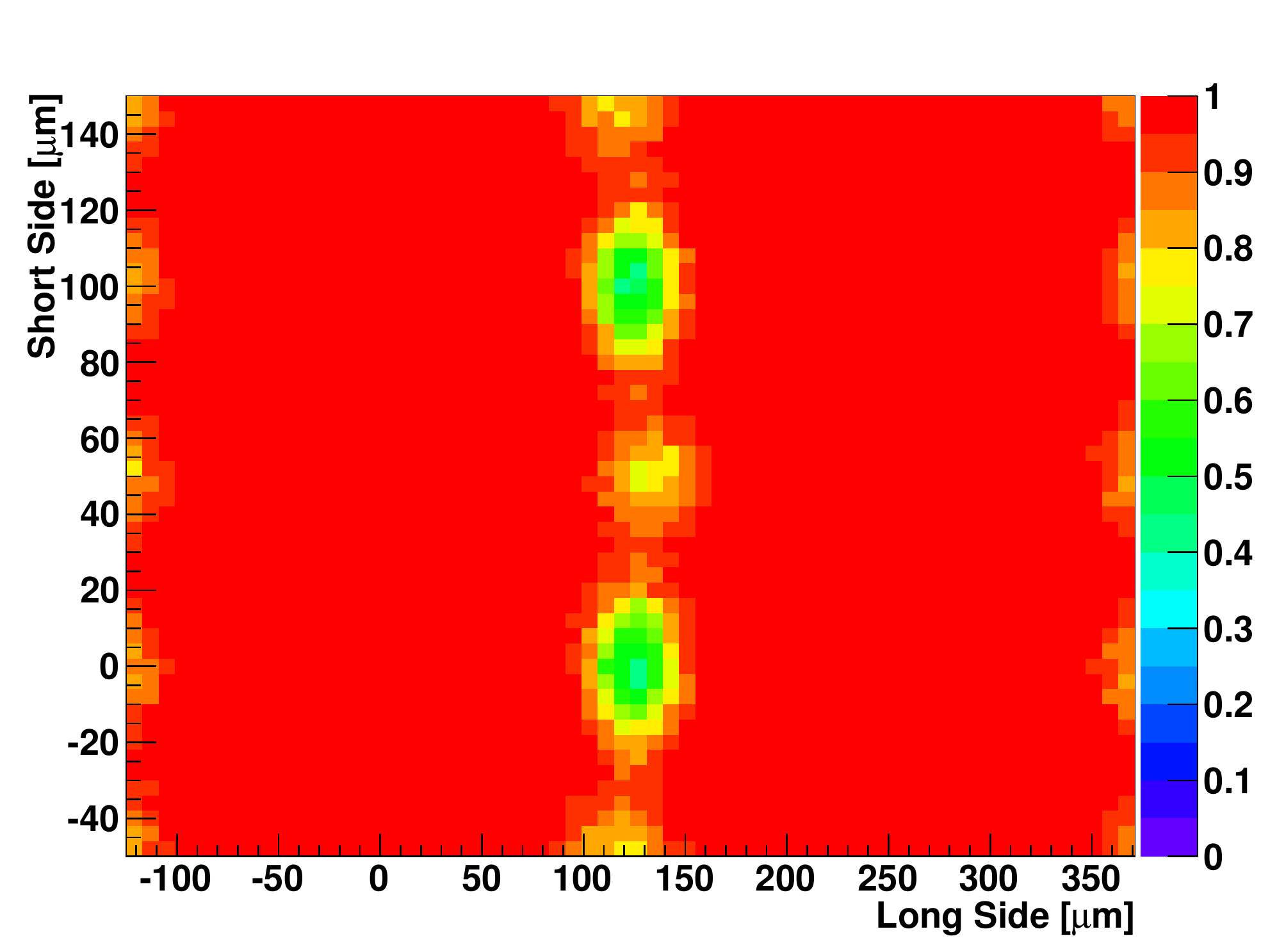}
\caption{\label{fig:annealingsoi3} In-pixel hit-efficiency of the $100\,\mu\mathrm{m}$ thick SOI sensor from HLL after an irradiation to $5\times10^{15}\,$n$_\text{eq.}/$cm$^2$ before (left) and after (right) 6 weeks of annealing at RT at $600\,$V. The efficiency map shows the area of $4\times2$ pixels. The efficiency before ($97.1\%$) and after ($97.0\%$) annealing are compatible.}
\end{figure}

In the present status of this study, two modules are investigated. The first one, assembled with a $100\,\mu\mathrm{m}$ thick sensor from an HLL production, was irradiated to a fluence of $5\times10^{15}\,$n$_\text{eq.}/$cm$^2$ with protons at the irradiation facility of the University of Birmingham \cite{birmingham}. It was measured before and after an annealing period of 6 weeks at RT. The hit efficiency of both measurements is shown in Fig. \ref{fig:annealingsoi3} for a bias voltage of $600\,$V. The received fluence is an upper limit for the possible situation of the third layer at the moment of the inner part replacement. Further measurements will be performed after 18 weeks and nearly one year to follow the evolution of the annealing. After 6 weeks of annealing no significant degradation of efficiency was found. The efficiency maps in Fig. \ref{fig:annealingsoi3} show the efficiency of the module given a specific position of the through-passing particle relative to the pixel cell. A matrix of $2\times4$ pixels is shown in each map and a clear drop of efficiency is evident in the corner region around each pixel. Every second corner is different from the others with an even more pronounced drop of efficiency due to the implemented punch-through structure. While the charge sharing between two pixels does not effect the efficiency, the charge sharing between four pixels causes the less pronounced drop of efficiency in the corners of the pixel cells. The efficiency drop due to the punch-through dot is reduced after annealing.

\begin{figure}[htbp]
\centering
\includegraphics[height=5.8cm,width=0.42\textwidth]{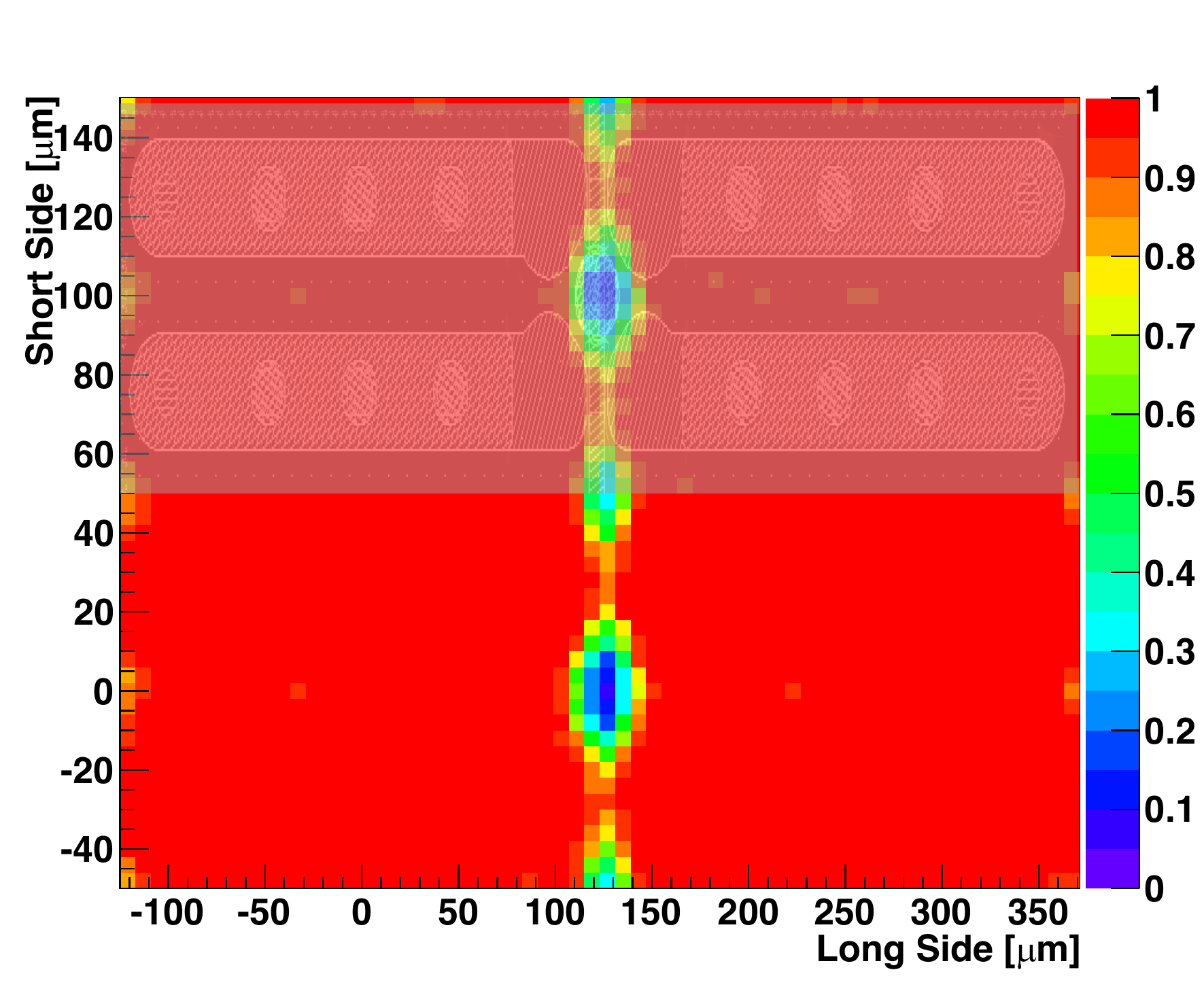}
\includegraphics[height=5.8cm,width=0.42\textwidth]{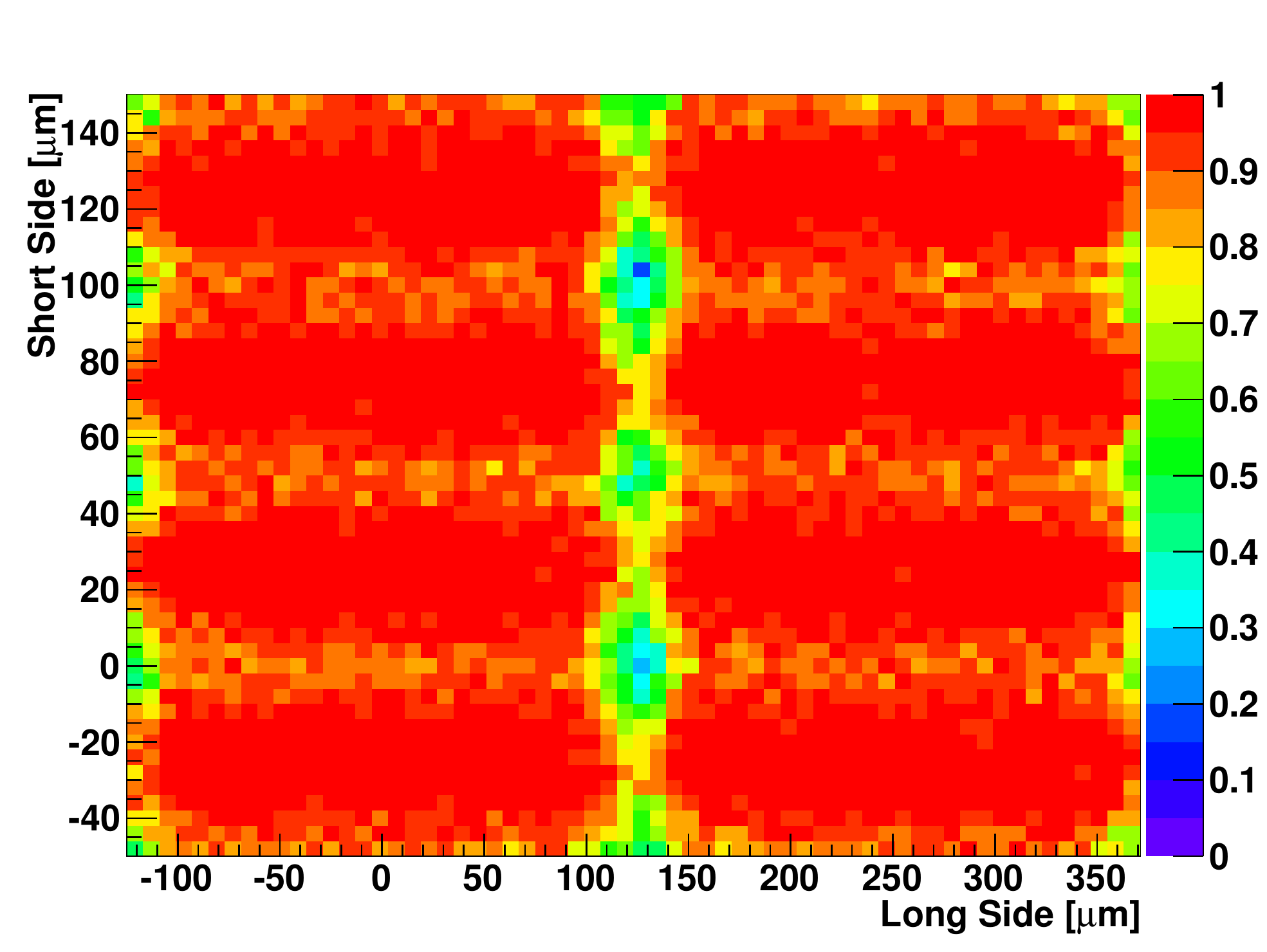}
\caption{\label{fig:annealingcis4} CiS sensor before (left) and after 100 hours of annealing at $60^\circ$C (right) - a 13 months duration scaled to RT. The module was irradiated to a fluence of $1\times10^{16}\,$n$_\text{eq.}/$cm$^2$. The efficiency before annealing is $96.6\%$ whereas it decreases to $92.2\%$ after annealing.}
\end{figure}

The second module, assembled with a $150\,\mu\mathrm{m}$ thick sensor produced at CiS, irradiated\footnote{at the CERN IRRAD facility using protons} to a fluence of $1\times10^{16}\,$n$_\text{eq.}/$cm$^2$ and annealed for 100 hours at around $60^\circ$C resulting in about 13 months of annealing at RT (scaling factor $\approx 100$) shows a significantly reduced efficiency of $92.2\%$ ($96.6\%$) after (before) annealing. The comparison of the in-pixel hit efficiency before and after annealing is shown in Fig. \ref{fig:annealingcis4}. This scenario points to the conditions at the end of lifetime of the ITk detector. Again, the efficiency is reduced in the corner region of the pixels which is due to charge sharing between four pixels and additionally due to charge losses associated to the punch-through dot in every other row. No efficiency losses between two pixels is visible in the not annealed case while a clear drop of efficiency in these cases is observed after annealing. On the contrary, the efficiency drop due to the punch-through dot is reduced after annealing, as it was observed for the first module. This effect is currently under investigation.

\begin{figure}[htbp]
\centering
\includegraphics[width=0.6\textwidth]{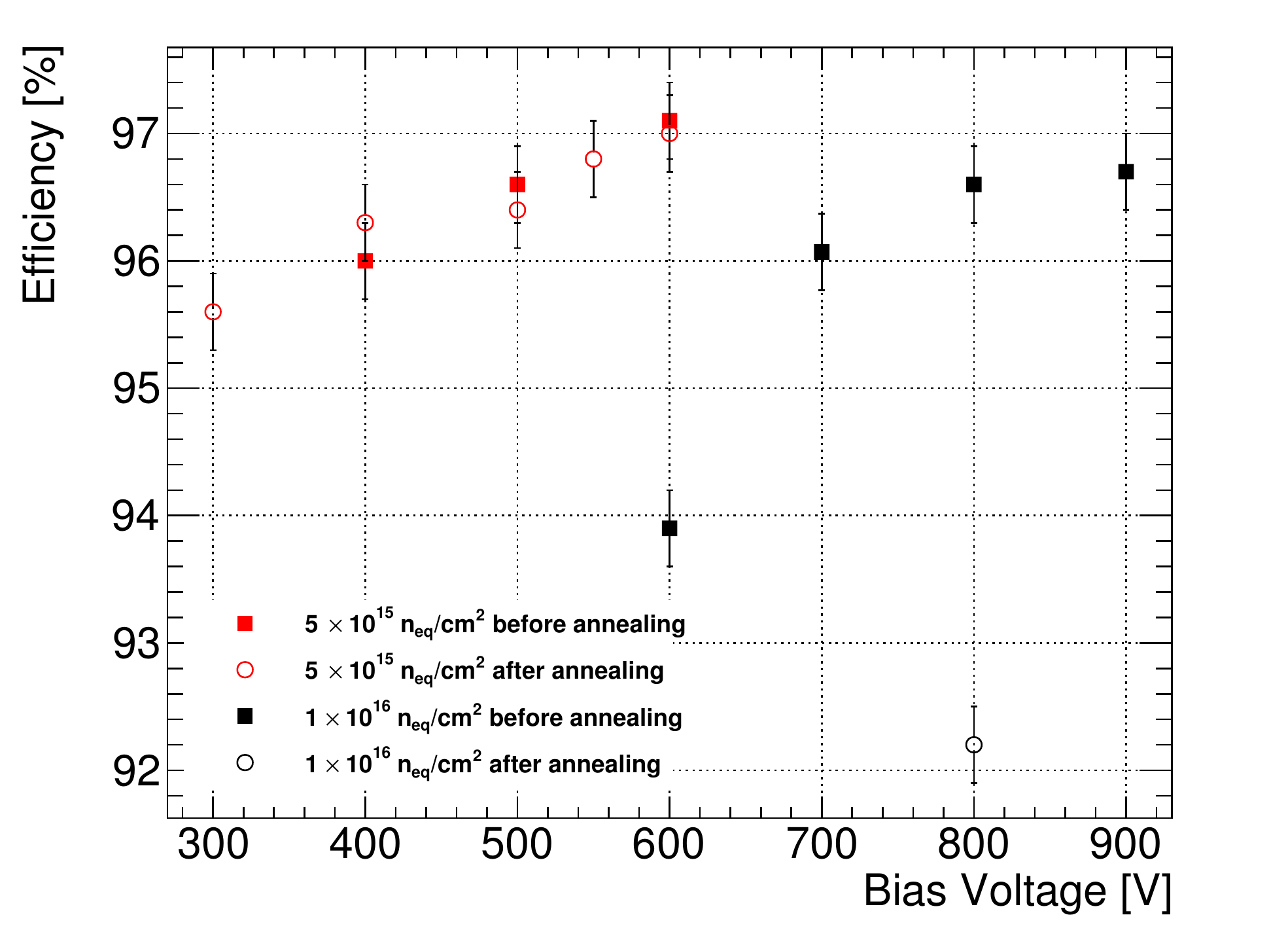}
\caption{\label{fig:annealingeffvsbias} Hit efficiency of the two measured modules before and after annealing as a function of bias voltage.}
\end{figure}

The efficiency of the two modules before and after annealing as a function of bias voltage is illustrated in Fig. \ref{fig:annealingeffvsbias}. While the efficiency is not effected by the annealing of 6 weeks at RT (SOI module), a clear decrease of efficiency was found in the case of 13 months annealing (CiS module).

%annealing has happened a lot in the past (ls1/2 and YETS) will also happen in future - especially exchange of layer 1/2, irr mdoules at UoB, pre irr vs 5 weeks vs high-T annealing TB studies - more to come

\section{Conclusion and outlook}

A 3D simulation of a $50\times50\,\mu\mathrm{m}^2$ pixel cell was carried out to identify possible process optimisations with respect to the p-spray properties. A clear improvement regarding the breakdown behaviour was found for lower p-spray doses and shallower p-spray implants. The same trend was observed in the measurement carried out on sensors of the latest HLL pixel sensor production of RD53 compatible devices. The simulations also indicate an increase of the breakdown voltage for larger pixel implants. Moreover, the pixel implant size was investigated regarding its influence on the charge collection efficiency. Small differences ($0.08-3.2\%$) were found between implant sizes of $35\,\mu\mathrm{m}$ and $45\,\mu\mathrm{m}$ before and after irradiation. The signal starts to degrade for a pixel implant size of $25\,\mu\mathrm{m}$ (up to $10.3\%$ less charge, compared to the best result of the respective irradiation level) and is significantly smaller for a pixel implant size of $10\,\mu\mathrm{m}$ (up to $30.4\%$ less charge). The effects of an increased inter-pixel capacity and the accompanying increasing noise will have to be considered to obtain the optimal signal to noise ratio.
\\ \\
The annealing study of irradiated modules shows that an annealing time of 6 weeks does not effect the hit efficiency of a $100\,\mu\mathrm{m}$ thick sensor at a fluence of $5\times10^{15}\,$n$_\text{eq.}/$cm$^2$. On the contrary, a CiS module irradiated to a fluence of $1\times10^{16}\,$n$_\text{eq.}/$cm$^2$ shows a significantly reduced efficiency of $92.2\%$ ($96.6\%$) after (before) annealing. The conclusion of this study can only be based on each module individually rather than a comparison of the two since the two modules have different thicknesses, different irradiation levels and different annealing times. More modules are in preparation at fluence levels of 1 and $2\times10^{15}\,$n$_\text{eq.}/$cm$^2$ which will also be measured after different annealing times, completing the preliminary picture of this study.

\acknowledgments

%Birmingham irr, Eutel, TBmon

This work has been partially performed in the framework of the CERN RD50 Collaboration. The authors thank L. Gonella for the irradiation of modules at the MC40 Cyclotron of the University of Birmingham and F.~Ravotti for the irradiation of modules at the CERN IRRAD facility at CERN-SPS. Supported by the H2020 project AIDA-2020, GA no. 654168. The authors thank the EUTelescope, TBmon2 and EUDAQ software developer teams. The authors thank J{\"o}rn Schwandt from the University of Hamburg and Rainer Richter and Peter Lechner from HLL for many fruitful discussions and helpful suggestions. 

% We suggest to always provide author, title and journal data:
% in short all the informations that clearly identify a document.

\end{document}